# Zeeman field-induced two-dimensional Weyl semimetal phase in cadmium arsenide

Binghao Guo[1], Wangqian Miao[1], Victor Huang[1], Alexander C. Lygo[1], Xi Dai[1,2], Susanne Stemmer[1,a]

[1] Materials Department, University of California, Santa Barbara, California 93106-5050, USA

[2] Department of Physics, Hong Kong University of Science and Technology, Hong Kong, China

[a] Corresponding author. Email: stemmer@mrl.ucsb.edu

**Abstract**

We report a topological phase transition in quantum-confined cadmium arsenide (Cd$_3$As$_2$) thin films under an in-plane Zeeman field when the Fermi level is tuned into the topological gap via an electric field. Symmetry considerations in this case predict the appearance of a two-dimensional Weyl semimetal (2D WSM), with a pair of Weyl nodes of opposite chirality at charge neutrality that are protected by space-time inversion ($C_2T$) symmetry. We show that the 2D WSM phase displays unique transport signatures, including saturated resistivities on the order of $h/e^2$ that persist over a range of in-plane magnetic fields. Moreover, applying a small out-of-plane magnetic field, while keeping the in-plane field within the stability range of the 2D WSM phase, gives rise to a well-developed odd integer quantum Hall effect, characteristic of degenerate, massive Weyl fermions. A minimal four-band $k \cdot p$ model of Cd$_3$As$_2$, which incorporates first-principles effective $g$ factors, qualitatively explains our findings.

Topologically non-trivial states of matter in two-dimensions (2D) have distinct advantages over their three-dimensional counterparts, as they can be readily manipulated by electric and/or magnetic fields, epitaxial strain, and proximity effects. A prime example is the quantum spin Hall insulator, known also as a 2D topological insulator (2D TI) [1,2], which supports helical edge modes inside a bulk energy gap in the presence of time-reversal symmetry. When subjected to in-plane magnetization, 2D TIs are predicted to give rise to an even richer variety of electronic states. These include quantum anomalous Hall insulators [3-6], density-wave states [7], as well as 2D (topological) semimetals [4,5,8]. While there exists a considerable amount of literature studying 2D TIs in out-of-plane magnetic fields (see, e.g., [9,10] and references cited therein), in-plane magnetization-induced phases have only recently been investigated [11-13]. As a result, the underlying nature of the observed transitions remains less clear. In principle, an in-plane magnetization can be supplied using magnetic dopants [5], or directly by an external magnetic field $B_{\text{ip}}$ that is oriented in the plane of the 2D system. The latter approach offers magnetic field tunability and is furthermore not limited to magnetic materials that require careful control of impurity concentrations and interactions. The main effect of $B_{\text{ip}}$ is to modify the electronic structure of 2D electronic systems via Zeeman coupling, $\Delta E_Z \sim g\mu_B B_{\text{ip}}$, where the $g$ factor is sample and material specific, and $\mu_B$ is the Bohr magneton.

In narrow gap semiconductors, a class that includes all experimentally observed 2D TIs, an in-plane Zeeman field may be sufficient to modify the topology of the Fermi surface. Several theoretical studies [4,8,14-16] have predicted that $B_{\text{ip}}$ can close the zero-field gap of a 2D TI and thereby drive the system into a metallic phase, when the hole- and electron-like subbands overlap in momentum space. Topological classification of this predicted (semi-)metallic phase remains ambiguous both experimentally and theoretically. Early transport data from inverted HgTe quantum wells suggested a conventional metal, based on observing a suppression of the local (and non-local) resistances in the diffusive transport regime [11].

In this Letter, we study the evolution of the recently reported [17] 2D TI phase of epitaxial cadmium arsenide ($Cd_3As_2$) thin films under in-plane and tilted magnetic fields, and identify the gapless phase as a 2D Weyl semimetal (WSM) with two valleys (nodes). A combined $C_2T$ symmetry protects the 2D WSM from opening an energy gap across the entire Brillouin zone, up to at least $B_{ip}$ ~ 14 T. Our conclusions are based on the following: (1) once the Fermi level is tuned to the charge neutrality point, closing of the inverted gap leads to a state with a saturated resistivity on the order of $h/e^2$, which spans a range of in-plane magnetic fields; (2) a well-developed odd integer quantum Hall effect appears when a small out-of-plane magnetic field is applied along with the in-plane field, a direct consequence of chiral zeroth Landau levels contributed by 2D Weyl nodes; and (3) experimental observations are consistent with a 4-band $k \cdot p$ model of confined $Cd_3As_2$ films [18,19] under an in-plane magnetic field and considering effective *g* factors for $Cd_3As_2$ thin films implemented within first-principles codes [20,21].

We begin by discussing expectations from symmetry considerations for a 2D TI $Cd_3As_2$ thin film in an in-plane magnetic field. In the absence of a magnetic field, $Cd_3As_2$ thin films possess 4/*mmm* point group and time reversal (*T*) symmetries [22]. Under an in-plane field, the symmetry reduces to the magnetic point group 2'/*m*', which contains the symmetry operators *E*, $C_2T$, $M_zT$, and inversion. As discussed in ref. [23], $C_2T$ symmetry eliminates one of the three Pauli matrices for the Hamiltonian at any specific *k* point in the 2D Brillouin zone. This leads to a 2D WSM within a finite phase region when the band gap is inverted by a tunable parameter (in the present case, the in-plane field strength). The significance of $C_2T$ symmetry to the local stability of the Weyl nodes against perturbations has also been discussed extensively in ref. [24]. Our computational results for the quantum well subbands confirm this symmetry analysis, as discussed next.

Figure 1(a) shows the dispersion in the $E$-$k_y$ plane computed for an 18 nm film at $B_{ip}$=10 T, obtained from a symmetry-invariant $k \cdot p$ Hamiltonian [25] for $Cd_3As_2$. The effective model parameters, and in particular the in-plane $g$ factors, are calculated by quasi-degenerate perturbation theory [26,27], implemented in a first-principle code [20,21]. As described in detail in [21] and in the Supplementary Materials [28], to obtain an accurate theoretical treatment of the magnitude of $B_{ip}$ effects, our approach involves further renormalizing the $g$ factors of $Cd_3As_2$ thin films, known to be large in the bulk (> 20 [32,33]), to account for the effects of quantum confinement. The dominating in-plane $g$ factor ($g_{1p}$, see [28]) is found to be ~12 in the bulk and ~ 13 for an 18-nm-thin film. The key result of this calculation is an isolated pair of Weyl nodes at the Fermi level, which are split along the direction perpendicular to the applied field. The results imply that their low-energy physics can be accessed at lab-scale magnetic fields. The degeneracy of the two Weyl nodes is furthermore guaranteed by bulk inversion symmetry present in $Cd_3As_2$ thin films [22,34,35]. Without a Zeeman field, the 18 nm film is a 2D TI, with doubly-degenerate subbands [Fig. 1(b)], consistent with our previous experimental results [17]. A calculated phase diagram of the band gap as a function of the in-plane field and the film thickness is shown in Fig. 1(c). The predicted thickness range for the 2D TI phase at zero field is in excellent agreement with our previous experiments [17]. We now turn to the experiments.

Transport measurements were carried out using top-gated Hall bars fabricated from high-mobility (001) $Cd_3As_2$ films, grown by molecular beam epitaxy to thicknesses of 18nm and 22 nm, respectively, on nearly lattice-matched buffer layers of $Al_{0.45}In_{0.55}Sb$, supported by (001) GaSb substrates [36]. Both film thicknesses fall within the 2D TI ("inverted") regime [17]. High-resolution x-ray reciprocal space maps, taken around the buffer 224 Bragg reflection, are shown in [28]. Data from four-point resistance measurements using low-frequency lock-in techniques are presented as two-dimensional resistivities or conductivities. All data were recorded at $T = 2$ K, unless stated otherwise. With no gate voltage applied, the 18 nm film had a low-field Hall mobility $\mu = 2.8 \times 10^4$ cm$^2$V$^{-1}$s$^{-1}$ at an electron density $n_{2D} = 3.6 \times 10^{11}$cm$^{-2}$, while the 22 nm film had $\mu = 1.7 \times 10^4$ cm$^2$V$^{-1}$s$^{-1}$ at $n_{2D} = 5.5 \times 10^{11}$cm$^{-2}$. For measurements with an additional out-of-plane field component $B_{oop}$, the total field $B_{tot}$ was fixed, while the samples were rotated *in situ* (for the angular alignment procedure, see ref. [28]). The angle $\theta$ relative to the sample normal (+z direction) was defined such that $\theta = 90°$ corresponds to the field positioned completely in plane. The fields in other cases are related to each other by $B_{oop} = B_{tot} \cos\theta$ and $B_{ip} = B_{tot} \sin\theta$.

Before we discuss the unusual nature of the in-plane results, we show, as an important point of comparison, the case with the magnetic field oriented fully out of plane. Figure 2(a) shows the longitudinal conductivity $\sigma_{xx}(B_{oop})$, measured on the 18 nm sample as a function of top-gate voltage ($V_g$). A modified gate voltage scale, $V_g - V_{CNP}$, is used to later aid comparison of films with slightly different as-grown carrier densities, where $V_{CNP}$ is the voltage corresponding to a global maximum in $\rho_{xx}(B_{tot} = 0)$. The evolution of subband Landau levels (LLs), and, in particular, the crossing of two $n = 0$ LLs, where $n$ is the LL index, is characteristic of the previously reported subband inversion in Cd$_3$As$_2$ [17]. A specific point to note is that spin or "isoparity" degeneracy (when spin is not a conserved quantum number [37]) is lifted at any finite $B_{oop}$ for $n > 0$ LLs [17,38], as evidenced by the appearance of completely developed, *even and odd* integer quantum Hall (QH) states at the same low $B_{oop}$ ~1.5 T. This point will become important later, because it clearly distinguishes the topological phases in $B_{oop}$ and $B_{ip}$. A higher-energy conduction subband contributes an additional set of LLs at 4 V, outside the low-energy window of interest for in-plane experiments, and we do not discuss it in this study.

Next, we examine the same device with the magnetic field fully in plane. In Fig. 2(b), we show the longitudinal resistivity $\rho_{xx}(B_{ip})$ of the 18 nm sample as a function of $V_g$, which tunes the 2D carrier system from $n$- to $p$-type transport regimes. Resistivity traces at constant $B_{ip}$ peak around 0 V, but the peak magnitude shows a non-monotonic evolution with increasing $B_{ip}$: the peak resistivity increases rapidly at low fields, followed by a sharp drop past ~1 T. The 22 nm $Cd_3As_2$ film, which lies at the other end of the thickness range for the inverted (2D TI) phase, demonstrates similar behavior, as shown in Fig. 2(c). Our results are qualitatively similar to that of inverted (001) HgTe quantum wells at comparable magnitudes of $B_{ip}$ [12]. Representative traces of $\rho_{xx}(V_g - V_{CNP})$ at fixed $B_{ip}$ are provided in [28].

In Fig. 3(a), we extract the peak resistivity under $B_{ip}$ from the two inverted samples to visualize the band gap closing under large Zeeman fields. The transition point, where $d\rho_{xx}/dB_{ip}$ goes to zero, is ~1 T in both samples. Beyond this transition point, the rate of gap closing is greater in the 18 nm sample than in the 22 nm sample. The gap is consequently closed at a higher field in the 22 nm sample. In the gapless region, shown in Fig. 3(b), both samples exhibit resistivities on the order of the resistance quantum $h/e^2$, with the 22 nm sample showing slightly lower values (~0.8 $h/e^2$). For the 18 nm sample, a > 90% reduction of resistivity as compared to zero-field values is observed, i.e., giant negative magnetoresistance. Also, while the 22 nm sample shows a saturated resistivity [Fig. 3(b)], the 18 nm sample has additional structure in the gapless region that is qualitatively reproducible between devices on the same film (hb1 vs. hb2) and that requires further investigation beyond the present study. We note that a resistivity value near $h/e^2$ is reasonably close to what is expected for 2D Dirac/Weyl systems [39,40]. The gap closing and transition to a semimetal phase is further confirmed by the temperature dependence of the resistivities, see [28]. The wide range of $B_{ip}$, within which the gapless phase is found, is one of the key experimental results, because it is consistent with the predicted wide stability range of the Zeeman field-induced 2D WSM phase, as discussed earlier.

To further characterize the nature of the 2D semimetal, we focus on its QH effect when both in-plane and out-of-plane magnetic fields are present. This is accomplished by tilting the film ($\theta < 90°$), while keeping $B_{tot}$ constant. The out-of-plane component will gap the $C_2T$-symmetric Weyl nodes, thus allowing us to investigate the 2D WSM's QH effect. We study the 18 nm sample, which enters the gapless phase at lower $B_{ip}$. In Fig. 4, we present sets of $\sigma_{xx}(V_g)$ and $\sigma_{xy}(V_g)$ traces, one set for each of four tilt angles, 80°, 75°, 70°, and 65° from panels (a) to (d), at $B_{tot} = 14$ T. The corresponding $B_{oop}$ values are 2.43 T, 3.62 T, 4.79 T, and 5.92 T with an uncertainty of $\pm$ 0.05 T [28]. $B_{ip}$ is greater than 12.5 T in all cases so that the sample remains in the $B_{ip}$ range of the gapless phase.

We make two main observations regarding the QH effects seen in Fig. 4. First, using the fact that a $\sigma_{xy} = 0$ plateau ($v = 0$) is present in all tilt cases and connects to $v = \pm 1$ plateaus, we deduce that the two $0^{th}$ LLs are spin-resolved, otherwise their contribution to the filling factor sequence would be in increments of two (or higher if there are other degeneracies). This is also consistent with their dispersion in $B_{oop}$ up to 5.92 T, which causes the $\sigma_{xy} = 0$ plateau to continuously widen on the $V_g$ scale. We contrast the behavior in the gapless phase with the $B_{oop}$ LL spectrum shown in Fig. 2(a), where the $\sigma_{xy} = 0$ plateau narrows down until it vanishes near 5 T when the two $0^{th}$ LLs meet.

Second, the LLs with $|n| \geq 1$ give rise to an *odd integer* sequence of filling factors. At $B_{oop} \leq 3.62$ T, as shown in Fig. 4(a) and 4(b), in addition to $v = \pm 1$, $\sigma_{xy}$ plateaus at $v = 3$, 5, and 7 are most prominent, but additional minima in $\sigma_{xx}(V_g)$ appear at electron densities corresponding to $v = 9$ and $v = 11$. We therefore conclude that higher order LLs are 2-fold degenerate when $B_{oop}$ is small, a remarkable contrast to our findings when there is no in-plane field [Fig. 2(a)]. Together with the spin-resolved $0^{th}$ LLs discussed earlier, this odd-integer-only sequence suggests the existence of a pair of massive Weyl fermions, whose filling factors are expected to follow a $2(n + ½)$ sequence [41,42]. The two Chern insulators, one that develops from the 2D TI, and the other from the 2D WSM state, are thus easily distinguished at low out-of-plane fields. In [28] we provide another 32 sets of conductivity traces, acquired at lower $B_{tot}$ values (down to 10 T) and using the same 4 tilt angles, showing that the results in Fig. 4 continue to hold for the full range of the gapless phase.

The key features in the data can be explained within a minimal model for a gapped 2D Weyl semimetal under a Zeeman field. For a Weyl point in a single valley ($K^+$), the Hamiltonian under a quantizing magnetic field is well-studied and can be written as:

$$H_{K^+} = \begin{pmatrix} \Delta & v_F \Pi^\dagger \\ v_F \Pi & -\Delta \end{pmatrix} \tag{1}$$

$$H_{K^+} = \begin{pmatrix} g_z B_z & v_F[(\hbar k_x + eB_z y) - \hbar \nabla_y] \\ v_F[(\hbar k_x + eB_z y) + \hbar \nabla_y] & -g_z B_z \end{pmatrix} \tag{2}$$

where Zeeman coupling is included as a mass term $\Delta = g_z B_z = g_z B_{oop}$, $g_z$ is an out-of-plane $g$ factor, and $v_F$ is an isotropic Fermi velocity. The energy eigenvalue for the $n^+ = 0$ Landau level is directly obtained as $E(n^+ = 0) = g_z B_z$, and the higher order LLs ($|n^+| \geq 1$) are $E(|n^+| \geq 1) = \pm\sqrt{|n_+|\hbar^2 \omega_0^2 + \Delta^2}$, where $\omega_0 = \sqrt{2} v_F / l_B$ and $l_B = \sqrt{\hbar/(eB_{oop})}$ is the magnetic length. The valley with opposite chirality ($K^-$) shares the same solutions for higher order LLs, but the $n^- = 0$ Landau level experiences a sign change, with an eigenvalue of $E(n^- = $ $0) = -g_z B_z$. This model of the 2D WSM readily explains the observed 2-fold degeneracy of higher order LLs as coming from an expected valley degree of freedom, while the absence of any degeneracy for the two $0^{th}$ LLs are accounted for by their chirality. By only keeping up to linear terms in momentum, however, the model cannot describe the lifting of the valley degeneracy at large $B_{oop}$, seen in Figs. 4(c) and 4(d), which creates additional $\sigma_{xy}$ plateaus at even filling factors of 2, 4, and 6. We note that the lifting of the degeneracy is not likely due to a Lifshitz transition at high $V_g$, since all data in Fig. 4 span the same $V_g$ range. This observation should motivate future theoretical work towards a more complete description. As in conventional semiconductor systems, such as AlAs quantum wells, lifting of valley degeneracy may originate from electron correlation effects [43,44].

To conclude, we comment on the observation of such a well-developed odd integer QH effect in a topological material. Odd integer QH sequences have long been sought after in three-dimensional (3D) TIs subjected to perpendicular magnetic fields [45-50], because they provide a clear transport signature of the Dirac fermions on their surfaces. In 3D TIs, this odd integer QH effect is typically thwarted by the energy mismatch between the 2D Dirac fermions on the top and bottom surfaces and the continued conduction of the side surfaces in a perpendicular magnetic field [51].

Finally, we note that the ideal 2D WSM reported here is particularly noteworthy, given the rarity of ideal WSMs in 3D [34]. Moreover, as discussed in this Letter, the route reported here is general and applies to materials beyond $Cd_3As_2$. The realization of model 2D WSMs may open up many opportunities, including quantized anomalous Hall effects [4] or by serving as a platform for engineering unconventional superconductivity.

## Acknowledgements

The authors thank S. Sun and Y. Chen for discussions. This work was primarily supported by the Air Force Office of Scientific Research (Grant Nos. FA9550-21-1-0180 and FA9550-22-1-0270). Part of this work made use of the MRL Shared Experimental Facilities, which are supported by the MRSEC Program of the U.S. NSF (Award No. DMR 1720256, and computational facilities supported by the U.S. NSF (CNS-1725797) and administered by the Center for Scientific Computing. B.G. acknowledges support from the Graduate Research Fellowship Program of the U.S. NSF (Grant No. 2139319), and the UCSB Quantum Foundry, which is funded via the Q-AMASE-i program of the U.S. NSF (Grant No. DMR-1906325).

**Figure Captions**

**Figure 1:** 2D Weyl semimetal phase induced by a Zeeman field in confined $Cd_3As_2$. (a) Dispersion of the semimetal phase with two Weyl nodes located at the Fermi level ($E = 0$ meV, dashed line). The 4 bands are spin resolved. For the $k \cdot p$ slab calculation, $Cd_3As_2$ has a thickness $d = 18$ nm along [001] with an in-plane field $B_{ip} = 10$ T. (b) Dispersion of the topological insulator phase with no external magnetic field. The spin-degenerate bands are drawn as dashed red lines. (b) Phase diagram of the $\Gamma$-point energy gap in the $B_{ip}$-$d$ parameter space. The thickness range corresponds to the band inversion regime. Blue triangle: 2D WSM. Red square: 2D TI.

**Figure 2:** Out-of-plane and in-plane magnetotransport. (a) Longitudinal conductivity $\sigma_{xx}$ as a function of the out-of-plane field $B_{oop}$ and the scaled (relative to $V_{CNP}$) top-gate voltage for the 18 nm sample (device: hb1). Quantum Hall plateau regions are indexed by their filling factor $\nu$ up to 6. The measured voltage $V_g$ is offset by $V_{CNP}$, which corresponds to the charge neutrality condition defined in the main text. (b) Longitudinal resistivity $\rho_{xx}$ as a function of the in-plane field $B_{ip}$ and the top-gate voltage for the 18 nm $Cd_3As_2$ sample (device: hb1). The same is shown in (c) for the 22 nm sample. $V_{CNP} = -1.35$ V for the 18 nm sample, $V_{CNP} = -1.165$ V for the 22 nm sample. Measurements were done at $T = 2$ K in the transverse configuration.

**Figure 3**: Suppressed resistivity on the order of $h/e^2$ in the WSM phase. (a) Extracted $\rho_{xx}$ at the charge neutrality condition as a function of $B_{ip}$ for the 18 nm sample (devices hb1, hb2 are on the same sample) and the 22 nm sample. (b) Close-up of the $B_{ip} > 9$ T region for the data in (a). Shading indicates a ± 20% interval around the resistance quantum $h/e^2 \approx 25.81$ k$\Omega$ (dashed line).

**Figure 4**: Odd integer quantum Hall effect and valley splitting. (a) Longitudinal and Hall conductivities, $\sigma_{xx}$ and $\sigma_{xy}$, as a function top-gate voltage for a fixed $B_{oop} = 2.43$ T. The same is shown in (b) for $B_{oop} = 3.62$ T, in (c) for $B_{oop} = 4.79$ T, and in (d) for $B_{oop} = 5.92$ T. The total magnetic field $B_{tot} = 14$ T, which is fixed, and $B_{ip} > 12.5$ T in the 4 cases. $\sigma_{xy}$ is drawn with different colors and $\sigma_{xx}$ is drawn in black. Dashed black lines, $v =$ odd integers. Bold lines, $v =$ even integers, which are also labelled. The $\sigma_{xy} = 0$ plateaus are labeled separately.

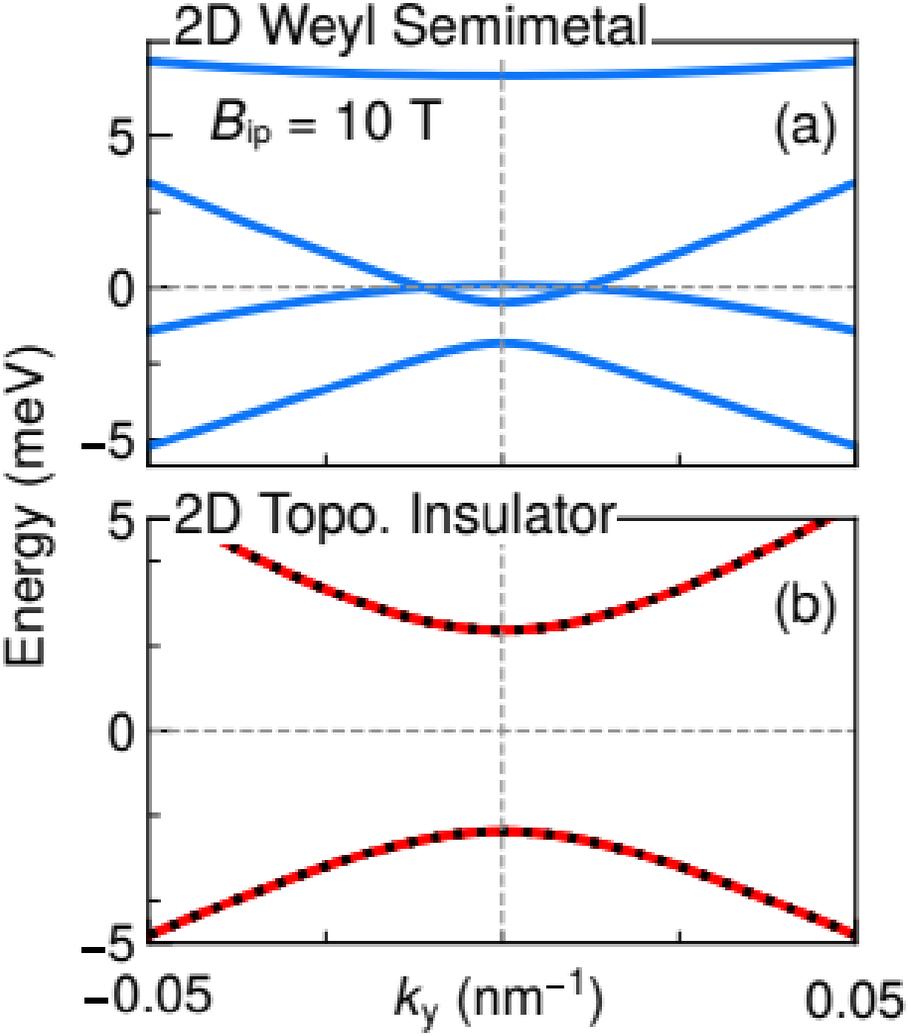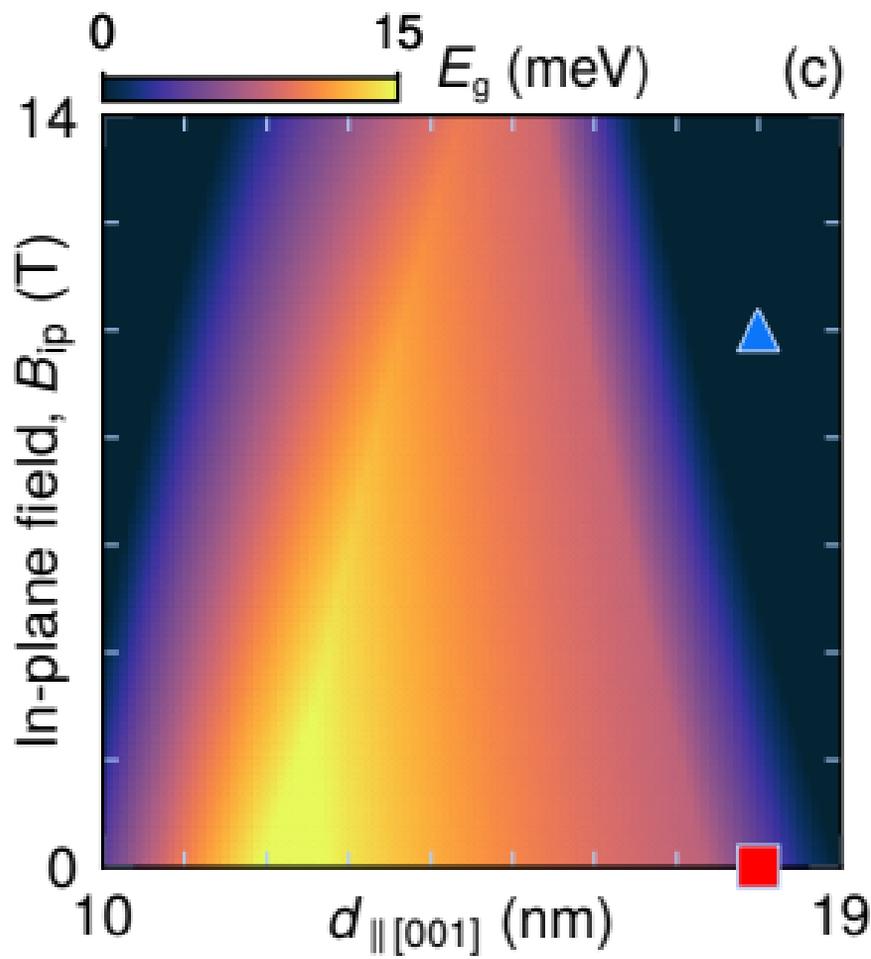

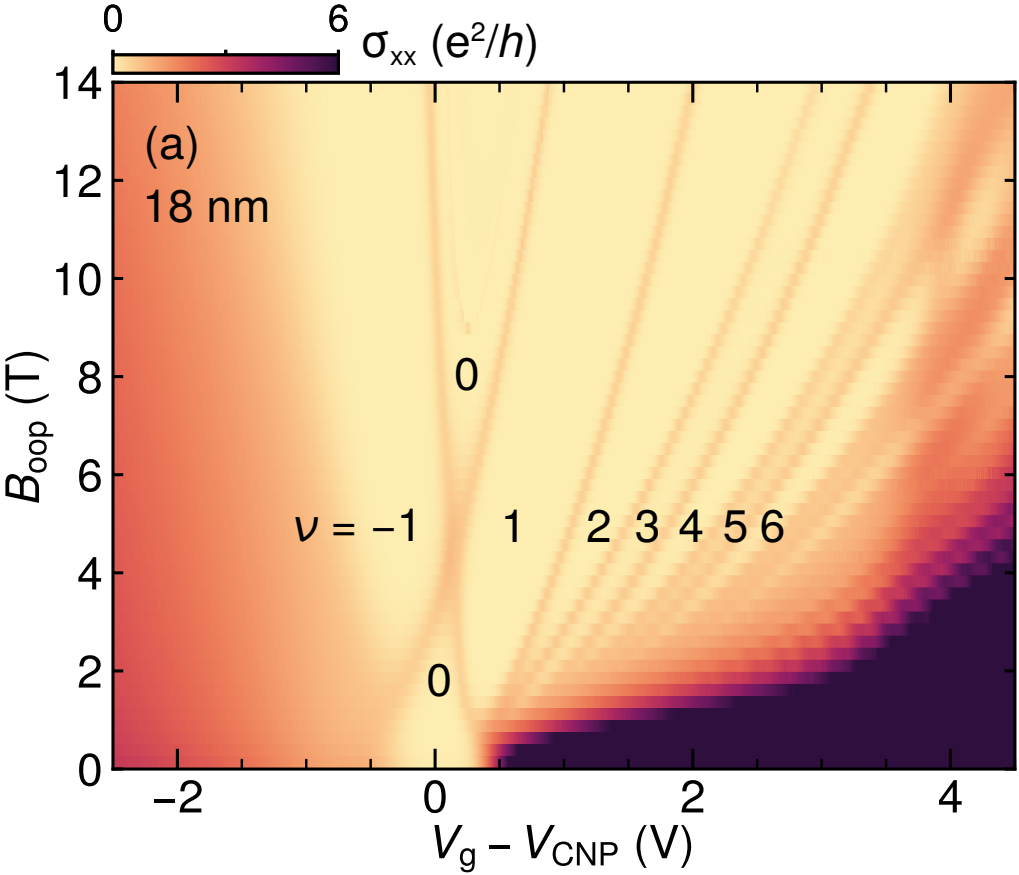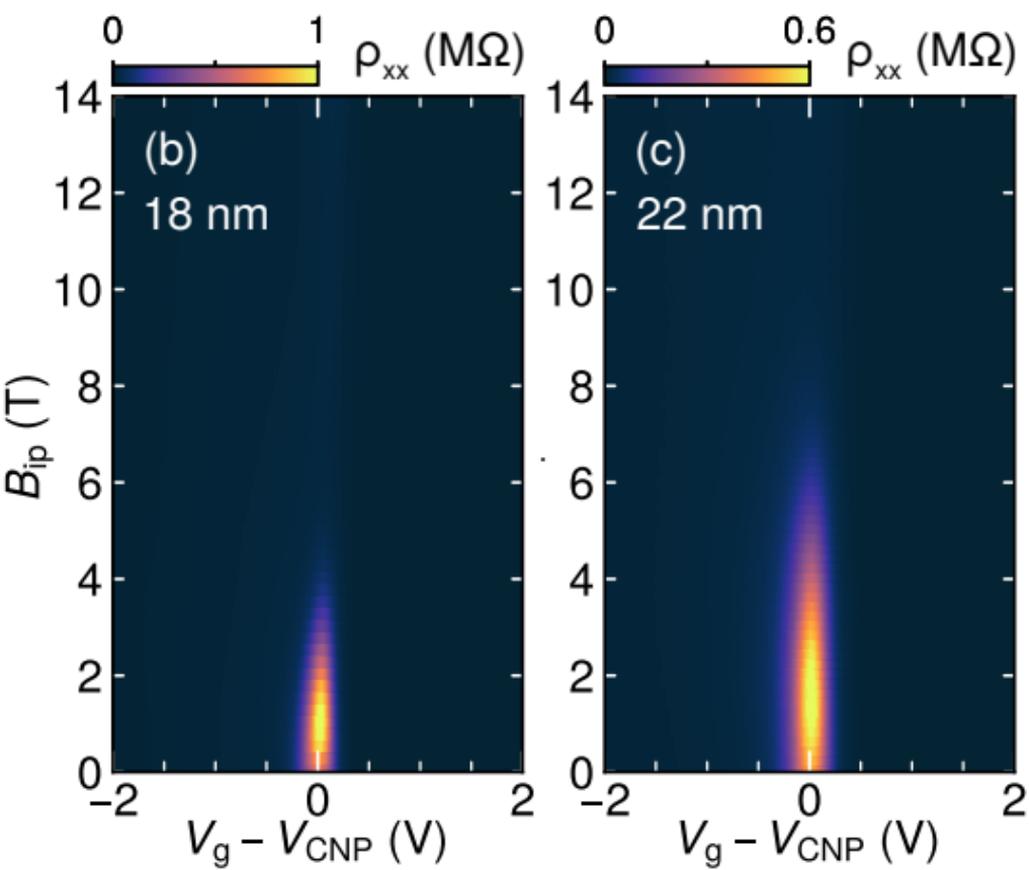

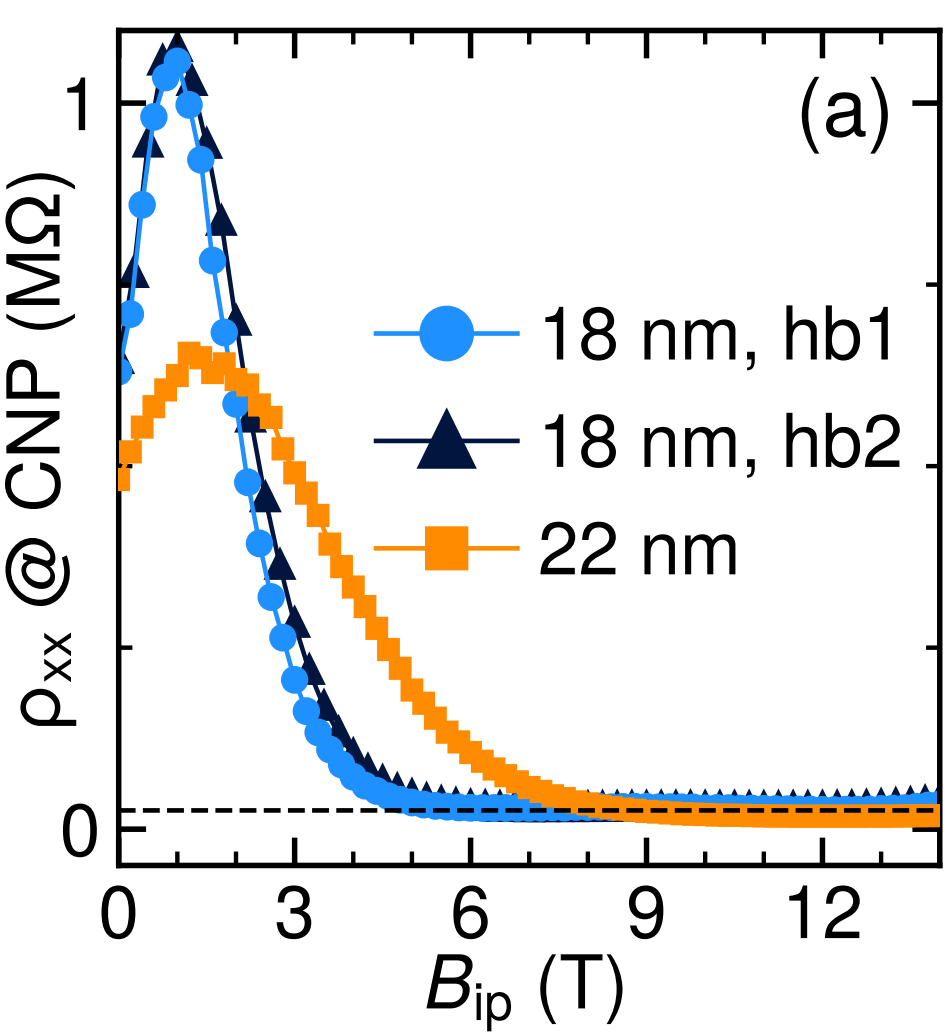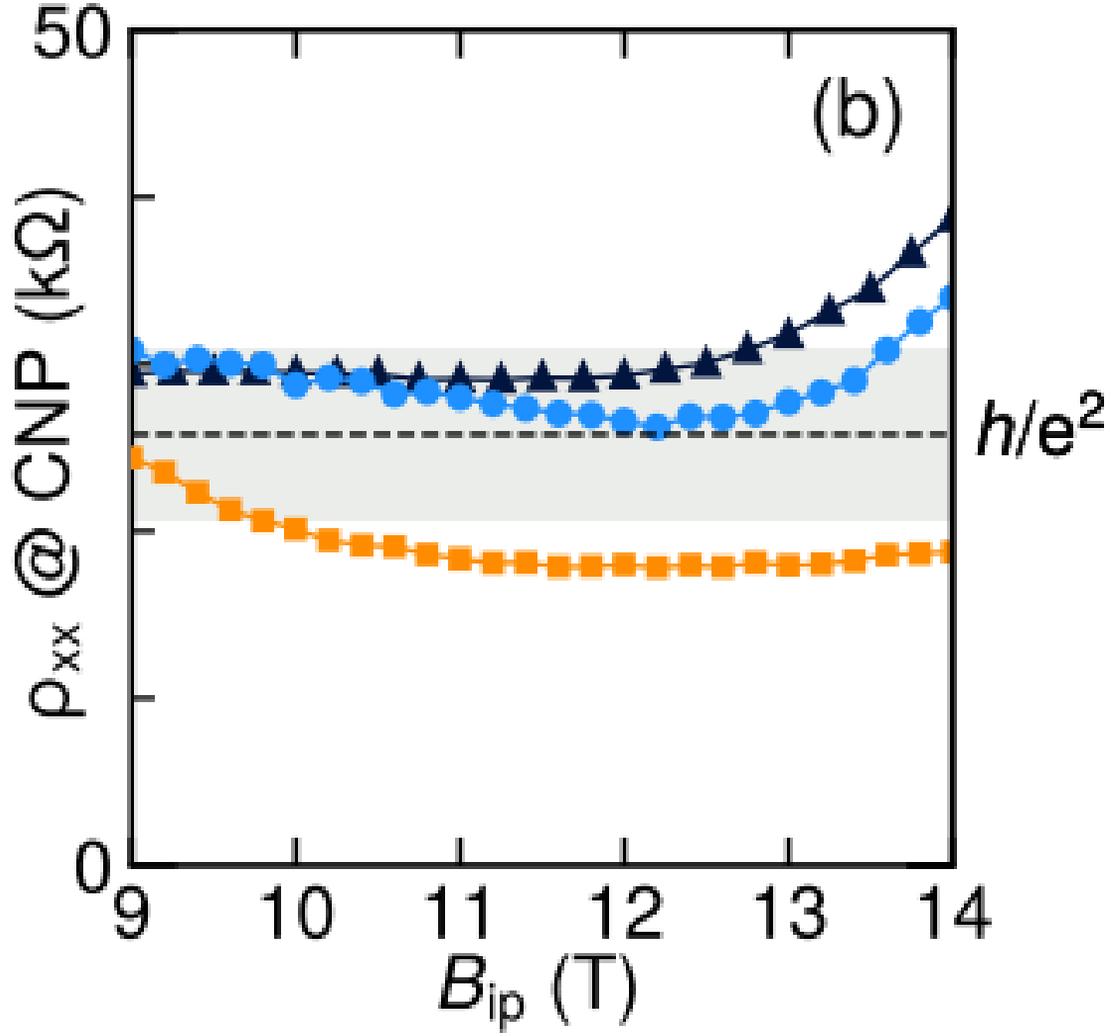

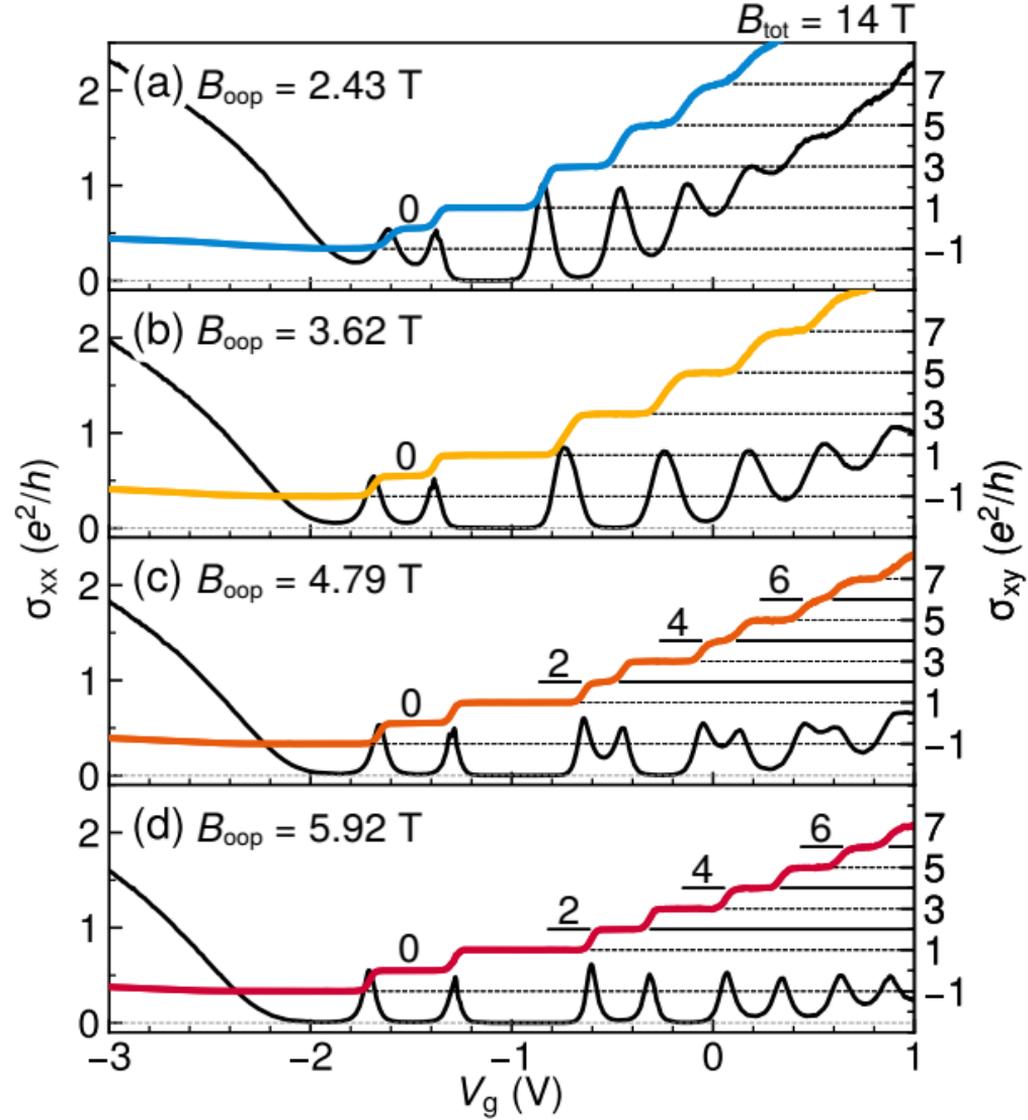

# Supplementary Material for

# Zeeman field-induced two-dimensional Weyl semimetal phase in cadmium arsenide

Binghao Guo[1], Wangqian Miao[1], Victor Huang[1], Alexander Lygo[1], Xi Dai[1,2], Susanne Stemmer[1]

[1]Materials Department, University of California, Santa Barbara, California, USA
[2]Department of Physics, The Hong Kong University of Science and Technology, Hong Kong, China

### $k \cdot p$ model for Cd$_3$As$_2$ & in-plane $g$ factor calculation

The crystal structure of bulk Cd$_3$As$_2$ has 4/mmm ($D_{4h}$) point group symmetry [1]. The effective orbitals near the Fermi surface can be represented as $\left|\frac{1}{2}, \pm\frac{1}{2}\right\rangle$ and $\left|\frac{3}{2}, \pm\frac{3}{2}\right\rangle$. Using these basis states, the corresponding matrix representations of the symmetry generators $C_{4z}$, $C_{2x}$, time reversal ($\mathcal{T}$), and inversion ($\mathcal{P}$) can be expressed as the following:

$$C_{4z} = \begin{bmatrix} e^{-i\frac{\pi}{4}} & 0 & 0 & 0 \\ 0 & e^{i\frac{\pi}{4}} & 0 & 0 \\ 0 & 0 & e^{-i\frac{3\pi}{4}} & 0 \\ 0 & 0 & 0 & e^{i\frac{3\pi}{4}} \end{bmatrix}, \quad C_{2x} = \begin{bmatrix} 0 & -i & 0 & 0 \\ -i & 0 & 0 & 0 \\ 0 & 0 & 0 & i \\ 0 & 0 & i & 0 \end{bmatrix}$$

$$\mathcal{T} = \begin{bmatrix} 0 & 1 & 0 & 0 \\ -1 & 0 & 0 & 0 \\ 0 & 0 & 0 & 1 \\ 0 & 0 & -1 & 0 \end{bmatrix} \mathcal{K}, \quad \mathcal{P} = \begin{bmatrix} 1 & 0 & 0 & 0 \\ 0 & 1 & 0 & 0 \\ 0 & 0 & -1 & 0 \\ 0 & 0 & 0 & -1 \end{bmatrix}.$$

The generators for a 4 × 4 matrix, denoted as $\sigma_i s_j$, can be classified according to the different representations of the $D_{4h}$ point group. Here, $\sigma_i$ refers to the Pauli matrix for the orbital part, while $s_j$ denotes the spin part (see Table 1 for details). Using these representations, we can construct the symmetry allowed $k \cdot p$ Hamiltonian $\mathcal{H}(\mathbf{k})$ by combining matrices and polynomials that belong to the same representation. We only consider up to second-order terms in $\mathcal{H}(\mathbf{k})$, which is sufficient for capturing the low-energy physics near the Fermi surface. The $k \cdot p$ Hamiltonian reads:

$$\mathcal{H}(\mathbf{k}) = (C_0 + C_1(k_x^2 + k_y^2) + C_2 k_z^2)\sigma_0 s_0 + (M_0 + M_1(k_x^2 + k_y^2) + M_2 k_z^2)\sigma_z s_0 \\ + A(k_x \sigma_x s_z - k_y \sigma_y s_0). \tag{1}$$

Furthermore, the in-plane magnetic field induced Zeeman term for the bulk $k \cdot p$ model is:

$$\mathcal{H}_{\text{Zeeman}} = \mu_B \left[ \tilde{g}_{1p}(\sigma_0 s_x B_x + \sigma_0 s_y B_y) + \tilde{g}_{2p}(\sigma_z s_x B_x + \sigma_z s_y B_y) \right]. \tag{2}$$

We adopt the following parameters for the $k \cdot p$ model, which are obtained from density functional theory (DFT) calculations: $C_0$ = -0.0145 eV, $C_1$ = 10.59 eV Å, $C_2$= 11.5 eV Å$^2$, $M_0$ = 0.0205 eV, $M_1$ = -18.77 eV Å$^2$, $M_2$ = -13.5 eV Å$^2$, $A$ = 0.889 eV Å, $g_{1p}$ = 11.6211, $g_{2p}$ = 0.5876, $\tilde{g}_{1p} = (g_{1p} + g_{2p})/2$, $\tilde{g}_{2p} = (g_{1p} - g_{2p})/2$. Here, $g_{1p}, g_{2p}$ are effective in-plane $g$-factors that we calculated using first-principles wavefunctions following the procedures described in [2]. For Cd$_3$As$_2$, the dominant $g$-factor is $g_{1p}$. In Fig. S1, we show the DFT band structure for bulk Cd$_3$As$_2$ compared to the one generated with our $k \cdot p$ model.

To simulate the quantum well (QW) structure of a Cd$_3$As$_2$ thin film grown along the [001] direction, we take the continuous limit where $k_z \rightarrow -i\partial_z$. The Hamiltonian $\mathcal{H}(\mathbf{k}) = \mathcal{H}(k_x, k_y, -i\partial_z)$ can then be expanded using a plane-wave basis $\psi_n(z) = \sqrt{\frac{2}{L}} \sin\left(\frac{n\pi z}{L} + \frac{n\pi}{2}\right)$, and $n$ is taken up to 50. The band structures for Cd$_3$As$_2$ QWs at three different thicknesses are shown in Fig. S1.

Using second order quasi-degenerate perturbation theory [3], the magnetic field induced Zeeman coupling is a gauge invariant term [2, 4, 5], and can be written as:

$$\mathcal{H}_{mm'}^{\text{Zeeman}} = \mu_B (\mathbf{g}_{mm'}^o + \mathbf{g}_{mm'}^s) \cdot \mathbf{B},$$
$$\mathbf{g}_{mm'}^o = \frac{-im_e}{2\hbar^2} \sum_{\ell,ijk} \left( \frac{1}{\epsilon_m - \epsilon_\ell} + \frac{1}{\epsilon_{m'} - \epsilon_\ell} \right) \pi_{ml} \pi_{lm'} \varepsilon_{ijk} \mathbf{e}_k. \quad (3)$$

where $\mu_B$ is the Bohr magnetron, $\mathbf{g}^s$ is the bare $g$-factor for the spin part, $\mathbf{g}^o$ is the renormalized $g$-factor after considering orbital effects from remote bands. $m, m'$ are band indices for the low energy subspace, where we preserve 4 bands near the Fermi level, and $\ell$ is a remote band index for the high energy subspace. The corresponding 4-band Hamiltonian for the QW shares the same structure as the Bernevig-Hughes-Zhang (BHZ) model, written as:

$$\mathcal{H}(\mathbf{k}) = (B_0 + B_1(k_x^2 + k_y^2))\sigma_0 s_0 + (D_0 + D_1(k_x^2 + k_y^2))\sigma_z s_0 + E(k_x \sigma_x s_z - k_y \sigma_y s_0). \quad (4)$$

The in-plane Zeeman field was described in Eq.(2) and the corresponding effective $g$-factors can be further enhanced in QW structures compared with their bulk values [6].

### 2D Weyl semimetal phase in Cd$_3$As$_2$ thin films

When an in-plane magnetic field of 10 T is applied, two Weyl points appear for an 18 nm Cd$_3$As$_2$ thin film. The pair of Weyl points are located along the direction perpendicular to the magnetic field, as shown in Fig.S2. In the following, we construct a Hamiltonian expanded to the linear order near the Weyl point for a single valley. It takes the form:

$$H^+ = v_x q_x \sigma_x + v_y q_y \sigma_y, \quad (5)$$

and is locally protected from gap opening by $C_2\mathcal{T}$ symmetry. Weyl points appear in pairs of opposite chirality, and are related by inversion symmetry in this case. The Weyl point in the opposite valley can therefore be expressed as:

$$H^- = -v_x q_x \sigma_x - v_y q_y \sigma_y. \quad (6)$$

Here, the key parameters to describe the Weyl points are their Fermi velocities, which we find to be $v_x = 0.48$ eV $\cdot$ Å ($7.3 \times 10^5$ m/s) and $v_y = 0.52$ eV $\cdot$ Å ($7.9 \times 10^5$ m/s).

**Thin film lattice parameter measurements**

Standard thin film XRD techniques were employed to measure the room temperature, in-plane lattice parameter of the alloy buffer layer, which is kept approximately constant over the $Cd_3As_2$ thickness series to minimize the influence of buffer layer-induced strain variation in $Cd_3As_2$. We are assuming here that the $Cd_3As_2$ epilayer remains semi-coherently strained to the alloy, including at $T \sim 2\,K$, the temperature at which transport measurements were made. The in-plane, pseudo-cubic lattice parameters of $Cd_3As_2$ were thus found to be 6.29 Å (18 nm) and 6.28 Å (22 nm), where the $Cd_3As_2$ film thickness is indicated in parentheses. Although we do not know the critical relaxation thickness of $Cd_3As_2$ on this specific alloy, a previous transmission electron diffraction study [7] indicates that 30 nm to 35 nm $Cd_3As_2$ films, thicker than all of the films studied here, remain strained to this type of buffer layer down to $T \sim 100\,K$ and supports the earlier assumption. $Cd_3As_2$ is also not known to undergo structural phase transitions below room temperature. Figure S3 shows reciprocal space maps (RSMs) of the region near the substrate GaSb 224 Bragg reflection. Data were acquired in an asymmetric geometry on a Rigaku SmartLab diffractometer using a HyPix3000 detector and Cu $K\alpha 1$ irradiation.

**Transport measurements & magnetic field alignment**

$Cd_3As_2$ Hall bars equipped with top gates were patterned by photolithography, with a detailed description provided in refs.[8, 9]. Two sizes of devices, with channel widths of 25 µm and 50 µm, were made, as shown in Fig. S4. We note here that $Ar^+$ ion milling was used to isolate mesas with a height $h \sim 110\,nm$, etching well past $Cd_3As_2$ into the alloy buffer layer, which is electrically insulating at measurement temperatures. Resistance measurements were done using low-frequency lock-in amplifier techniques at $T \approx 2\,K$. A 10 µV AC voltage (17.78 Hz) was applied across a 10.1 MΩ series resistor and we measured the source-drain current (which was <1 nA). This setup prevents self-heating effects that can be severe near charge neutrality in $Cd_3As_2$ thin films [8]. The resulting voltages, which were small (~ µV), were measured with SR560 voltage pre-amplifiers with nominal gains of either 100 or 500. Because of gate hysteresis from interface trap states, we are limited to analyzing data acquired using the same gate voltage sweep direction, i.e., from positive to negative [8].

For in-plane (and titled field) measurements, the samples were attached with varnish (on one edge) to a mechanical rotator stage, where a 90° index corresponds to a (nominally) completely in-plane position. Because all $Cd_3As_2$ films we studied here were grown on 1.5° miscut GaSb substrates, using the rotator was necessary to minimize alignment errors after mounting. We used our device effectively as a Hall sensor to find the angular index position that corresponded to a zero Hall signal, because a pure, in-plane field component should not, conventionally, generate a Hall voltage. For example, on the 18 nm $Cd_3As_2$ film, with the total magnetic field set to 14 T (to maximize the Hall signal) and the gate voltage set to 0 V (to position the Fermi level in an electron-like subband and not on a quantum Hall plateau), the rotator index was stepped from 100° to 80° (Fig. S4) and the Hall signal passed through zero at 85.5°, consistent with a misalignment of a few degrees caused by the miscut and mounting error. Given a step size of $\sim 0.17°$, we estimate the out-of-plane field error to be $\sim 0.05\,T$. Fig. S4 also shows that backlash is significant at 2 K when approaching the same index from opposite directions. After fixing the approach direction, the Hall signals converge and allow for meaningful angular comparisons. The approach speed is kept at 3 °/min for all measurements.

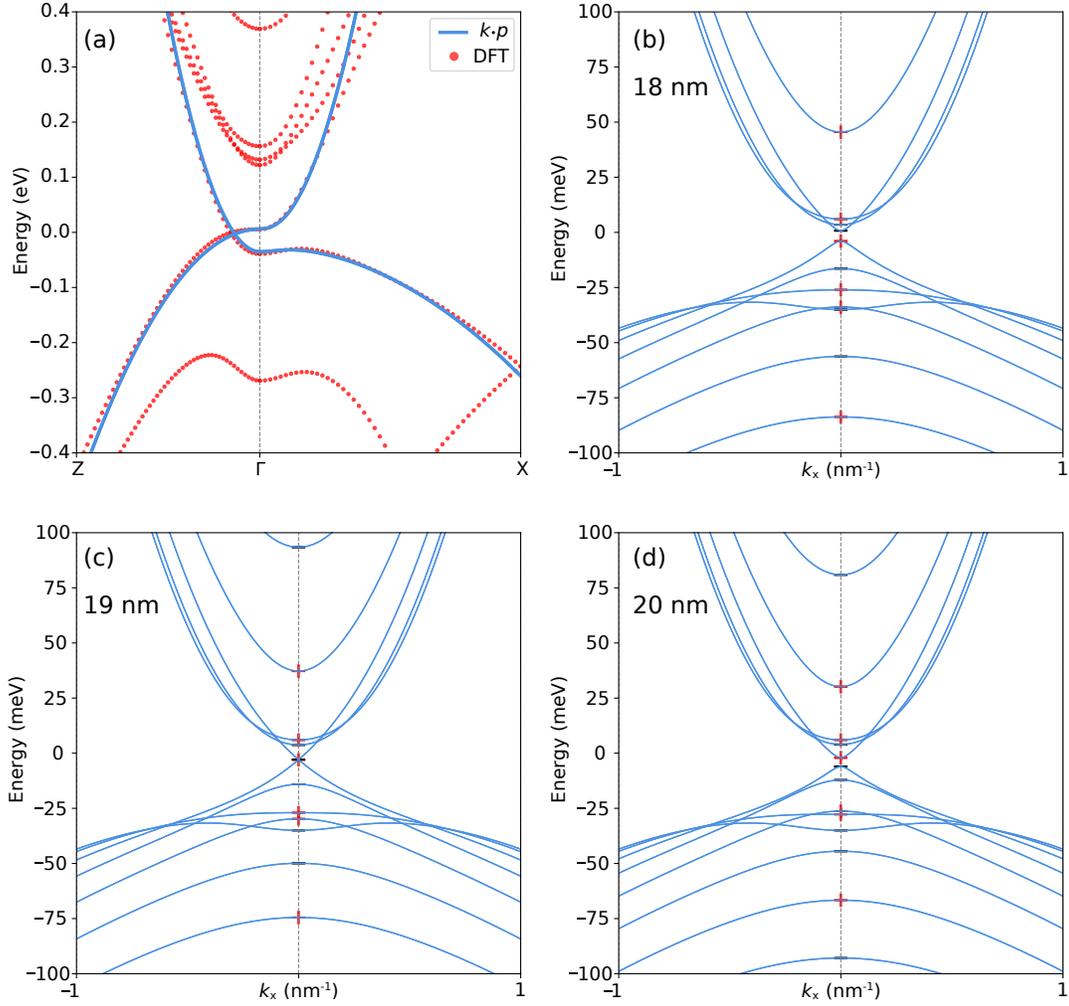

**Figure S1: (a)** Band structure (red dots) calculated from first-principles (DFT) along the high symmetry path $Z$-$\Gamma$-$X$ for bulk $Cd_3As_2$, compared with the one solved from the $k \cdot p$ Haimiltonian (blue lines). Good consistency is obtained near the Fermi level ($\pm$ 0.1 eV). The DFT calculation was carried out using the generalized gradient approximation (GGA) of the PBE functional with VASP. The planewave energy cutoff was set to 500 eV and a $6 \times 6 \times 6$ $k$-mesh was applied for the self consistent calculation. Our result is in close agreement with a recent DFT calculation [10]. **(b)-(d)** Band structures calculated for $Cd_3As_2$ slabs. **(b)** 18 nm $Cd_3As_2$ slab ($Z_2$ = 1). **(c)** 19 nm $Cd_3As_2$ slab (critical point). **(d)** 20 nm $Cd_3As_2$ slab ($Z_2$ = 0). The parity for each subband at the $\Gamma$ point is labeled. The corresponding $Z_2$ topological invariant was calculated using the Fu-Kane formula [11].

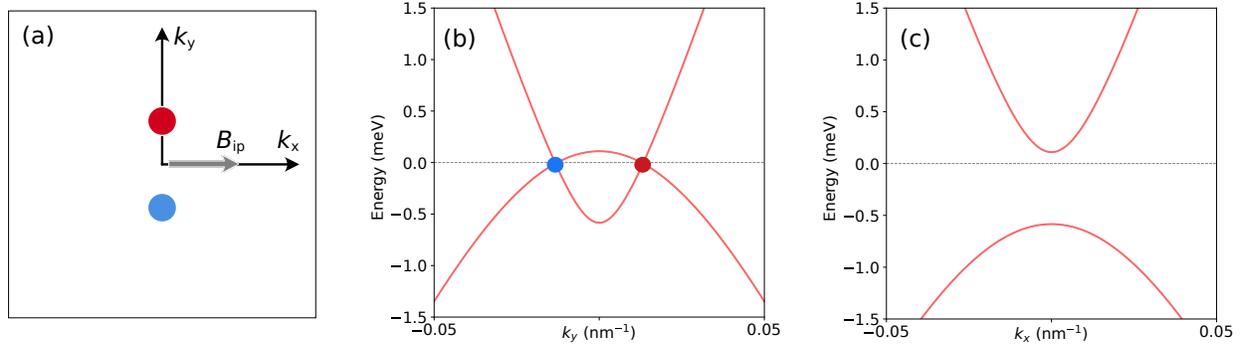

**Figure S2: (a)** Illustration showing the momentum space locations of the two Weyl points denoted as different colored dots. Weyl points are located along the $y$-axis which is perpendicular to the in-plane magnetic field. **(b)** Band structure of the 2D WSM along the $k_y$ axis, with the Fermi level indicated by a grey dashed line. This is the same data as shown in **Fig. 1** in the main text, but in a lower energy range. **(c)** Band structure of the 2D WSM along the $k_x$ axis.

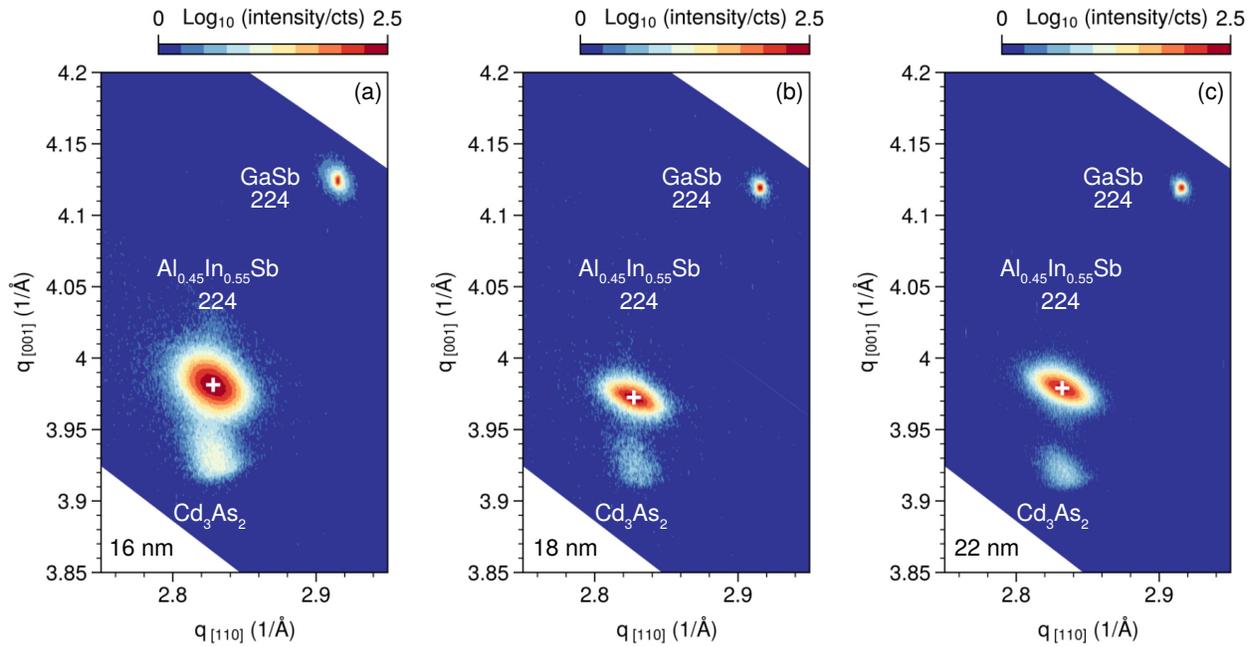

**Figure S3:** In-plane lattice parameter measurements. **(a)-(c)** Reciprocal space maps showing the epitaxial alignment of $Cd_3As_2$ with the buffer layer (+, white) and the substrate, for samples with varying $Cd_3As_2$ thickness. The $Al_{0.45}Ga_{0.55}Sb$ buffer layer and the $Cd_3As_2$ peaks are well-aligned in the vertical direction in all 3 samples, indicating that they share the same in-plane lattice parameter. The 18 nm and 22 nm films were studied in the main text. Directly extracting the peak positions of $Cd_3As_2$ reflections is challenging owning to their small scattering volume. They correspond to 4416 reflections using the notation for the conventional tetragonal cell of $Cd_3As_2$. Data were taken at room temperature and the definition of the scattering vector $q$ here includes a $2\pi$ factor.

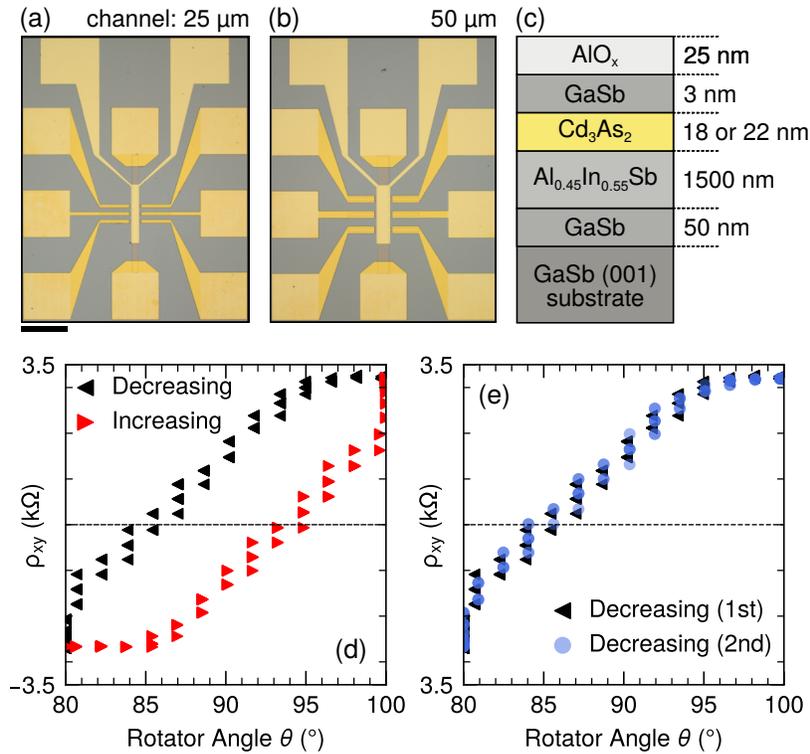

**Figure S4:** Sample overview and field alignment. **(a)** Optical image of a top-gated Hall bar with channel width of $25\,\mu m$, and **(b)** $50\,\mu m$. Scale bar, $100\,\mu m$. **(c)** Schematic of the $Cd_3As_2$ heterostructure cross-section, including the dielectric layer. **(d)** Hall resistivity as a function of rotator index position. The nominally in-plane condition at $90°$ produces an appreciable Hall signal, which can be zeroed out by decreasing the position to $85.5°$ (black triangles). This position can then be defined as completely in-plane. Approaching the same index positions from the opposite direction (red triangles) also produces different Hall signals (i.e., backlash). **(e)** Backlash can be minimized when the approach direction and the approach speed are kept the same. Data were acquired at $2\,K$ on the $18\,nm$ $Cd_3As_2$ sample (hb1) with $0\,V$ applied to the gate and a total magnetic field of $14\,T$.

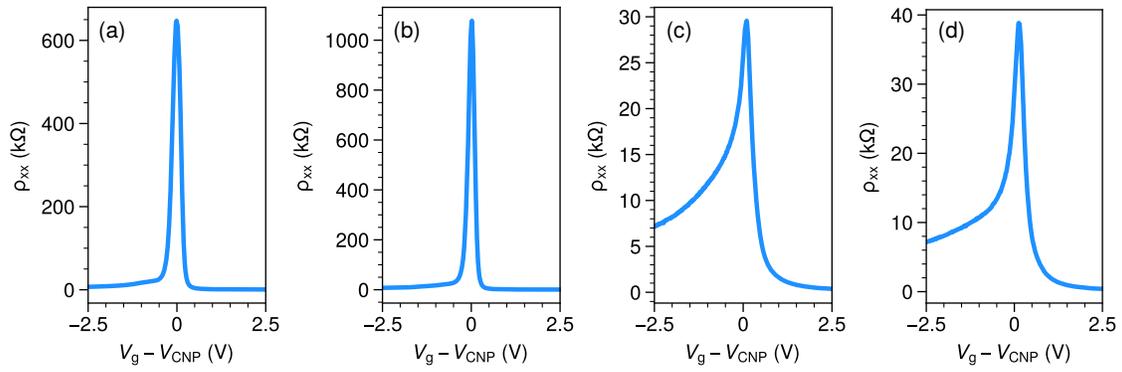

**Figure S5:** Representative $\rho_{xx}$ line cuts at constant in-plane magnetic fields for the $18\,\text{nm}$ sample (device: hb1). **(a)**-**(d)** The in-plane magnetic field is fixed in each panel to $0\,\text{T}$, $1\,\text{T}$, $10\,\text{T}$, $14\,\text{T}$, respectively. $V_{\text{CNP}} = -1.35\,\text{V}$. Data were acquired at $2\,\text{K}$.

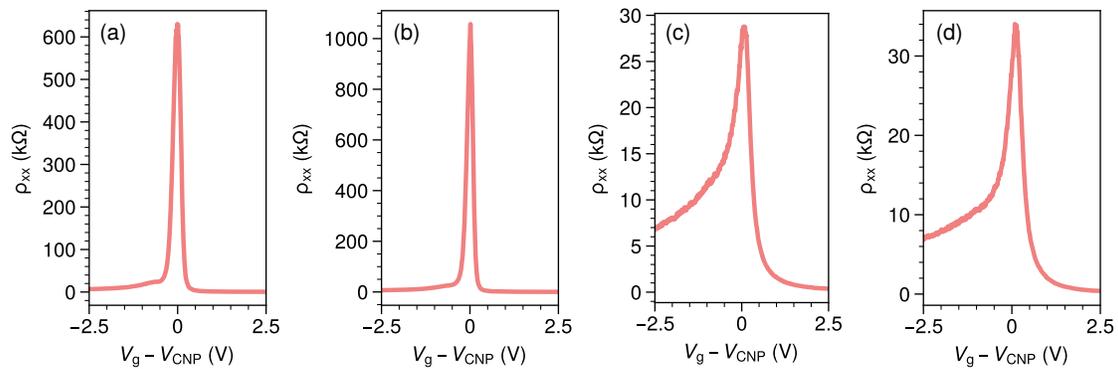

**Figure S6:** Representative $\rho_{xx}$ line cuts at constant in-plane magnetic fields for the 18 nm sample (device: hb2). **(a)**-**(d)** The in-plane magnetic field is fixed in each panel to 0 T, 1 T, 10 T, 14 T, respectively. $V_{\text{CNP}} = -1.37$ V. Data were acquired at 2 K.

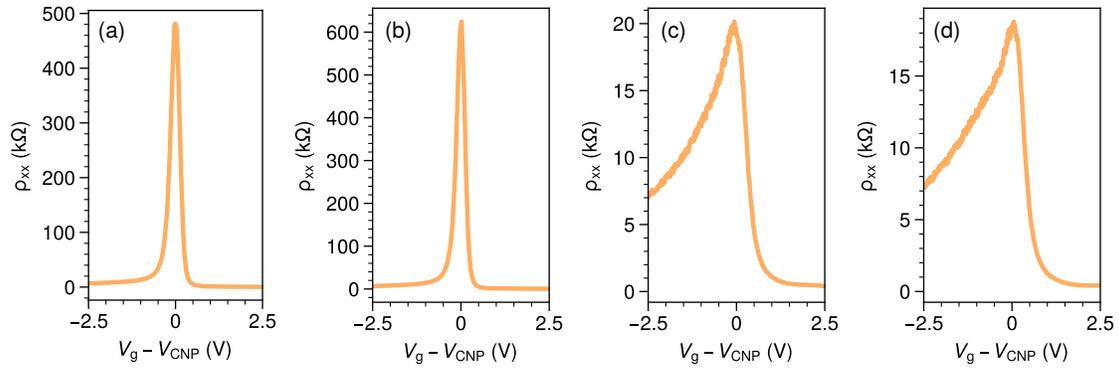

**Figure S7:** Representative $\rho_{xx}$ line cuts at constant in-plane magnetic fields for the $22\,\text{nm}$ sample. **(a)**-**(d)** The in-plane magnetic field is fixed in each panel to $0\,\text{T}$, $1\,\text{T}$, $10\,\text{T}$, $14\,\text{T}$, respectively. $V_{\text{CNP}} = -1.165\,\text{V}$. Data were acquired at $2\,\text{K}$.

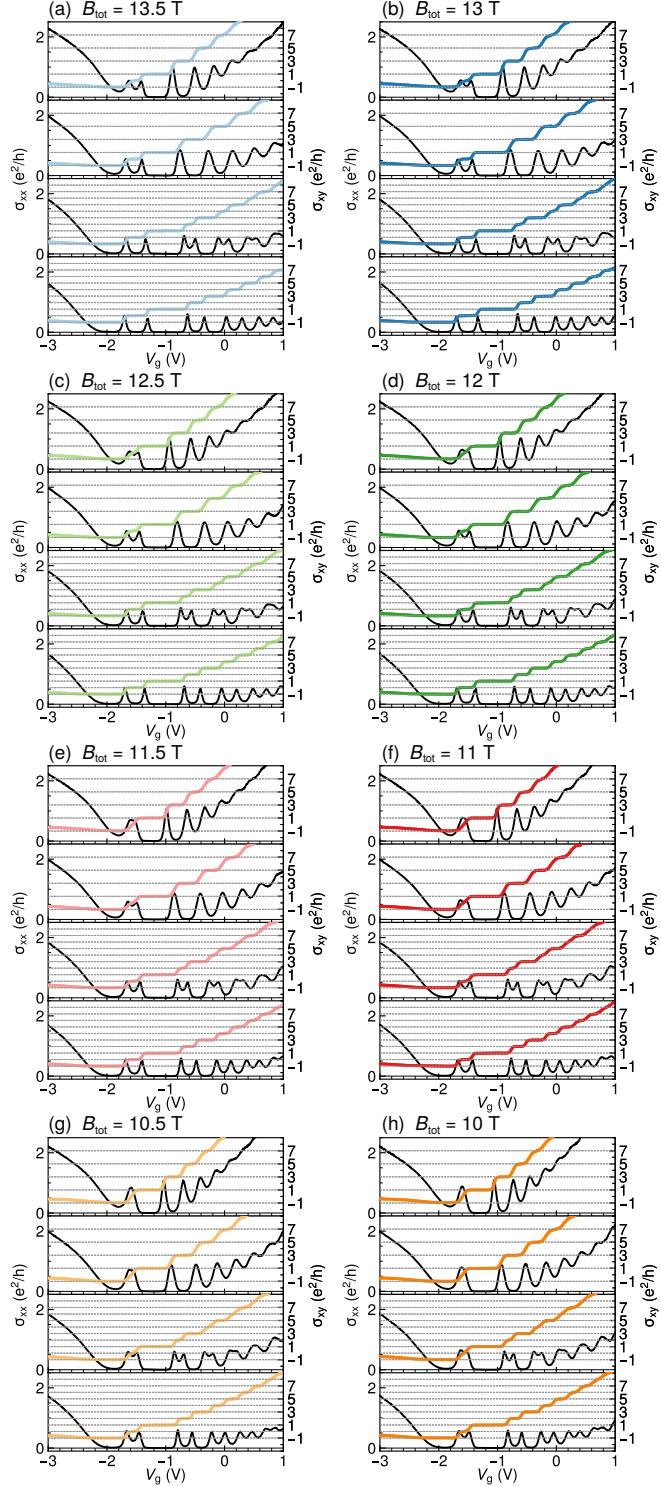

**Figure S8:** Additional $\sigma_{xy}$ and $\sigma_{xx}$ datasets from the 18 nm sample (hb1). **(a)**-**(h)** The total magnetic field is fixed in each panel and ranges from 13.5 T to 10 T in 0.5 T steps. Similar to **Figure 3** in the main text, the tilt angles for each panel are 80°, 75°, 70°, and 65° from top to bottom. Colored traces, $\sigma_{xy}$. Black traces, $\sigma_{xx}$. Each of the top two sub-panels have odd-integer filling factor plateaus indicated by dashed lines, while the bottom two sub-panels have all of the integers indicated. Data were acquired at 2 K.

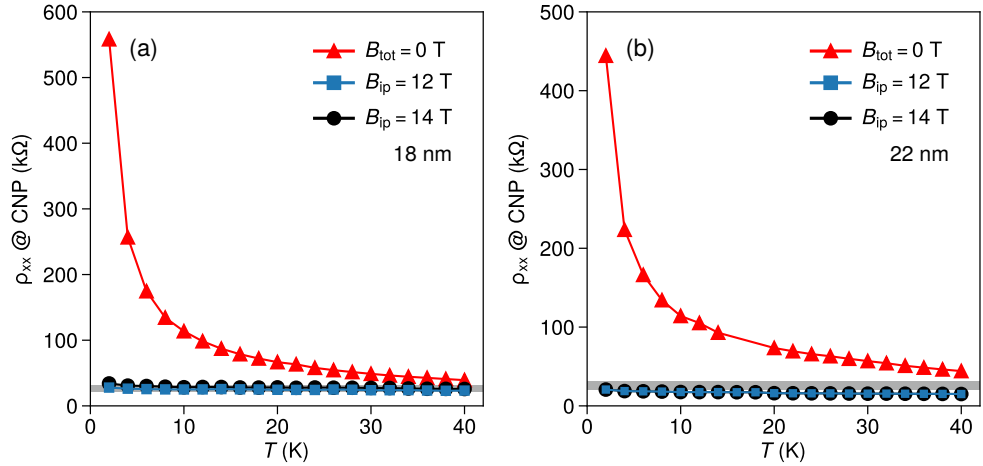

**Figure S9:** Temperature dependence of resistivity at the charge neutrality point. **(a)** Longitudinal resistivity ($\rho_{xx}$) at the charge neutrality point voltage ($V_{CNP}$) as a function of temperature for the 18 nm sample, with external magnetic fields of either $B_{tot} = 0$ T (red, triangle), $B_{ip} = 12$ T (blue, square), or $B_{ip} = 14$ T (black, circle). Shaded band indicates a ± 20% interval around the resistance quantum $h/e^2 \approx 25.81$ kΩ. **(b)** $\rho_{xx}$ at $V_{CNP}$ as a function of temperature for the 22 nm sample, with the same field values as in **(a)**.

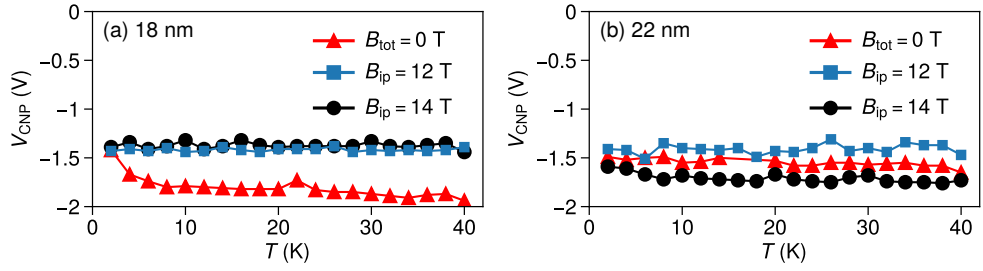

**Figure S10:** Temperature dependence of the charge neutrality point voltage. **(a)** $V_{\text{CNP}}$ as a function of temperature for the 18 nm sample, with external magnetic fields of either $B_{\text{tot}} = 0$ T, $B_{\text{ip}} = 12$ T, or $B_{\text{ip}} = 14$ T. **(b)** $V_{\text{CNP}}$ as a function of temperature for the 22 nm sample, with the same field values as in **(a)**.

**Table 1:** Invariant expansion for the $D_{4h}$ point group

| $D_{4h}$ | $\mathcal{T}$ | Matrices | Polynomials | $D_{4h}$ | $\mathcal{T}$ | Matrices | Polynomials |
|---|---|---|---|---|---|---|---|
| $A_{1g}$ | + | $\sigma_0 s_0, \sigma_z s_0$ | $1, k_x^2 + k_y^2, k_z^2$ | $B_{2u}$ | + | $\sigma_y s_x$ | |
| $A_{2g}$ | − | $\sigma_0 s_z, \sigma_z s_z$ | $B_z$ | $B_{2u}$ | − | $\sigma_x s_x$ | |
| $A_{2u}$ | − | | $k_z$ | $E_g$ | + | | $(k_x k_z, k_y k_z)$ |
| $B_{1g}$ | + | | $k_x^2 - k_y^2$ | $E_g$ | − | $(\sigma_0 s_x, \sigma_0 s_y), (\sigma_z s_x, \sigma_z s_y)$ | $(B_x, B_y)$ |
| $B_{1u}$ | + | $\sigma_y s_y$ | | $E_u$ | + | $(\sigma_x s_0, \sigma_y s_z)$ | |
| $B_{1u}$ | − | $\sigma_x s_y$ | | $E_u$ | − | $(\sigma_x s_z, -\sigma_y s_0)$ | $(k_x, k_y)$ |
| $B_{2g}$ | + | | $k_x, k_y$ | | | | |